\documentstyle[11pt]{article}
\newcommand{\be}{\begin{equation}}
\newcommand{\ee}{\end{equation}}
\newcommand{\ba}{\begin{eqnarray}}
\newcommand{\ea}{\end{eqnarray}}
\newcommand{\bann}{\begin{eqnarray*}}
\newcommand{\eann}{\end{eqnarray*}}

\begin{document}
\hbadness=10000
\setcounter{page}{1}

\title{Bose Einstein Correlations and Quantum Field Theory
\footnote{Invited talk at the Second German-Poöish
Symposium, Zakopane 1995, to appear in the volume "New Ideas
in the Theory of Fundamental Interactions", Acta Physica
Polonica B, 1996, editors H.D.Doebner, M.Pawlowski, and R.
Raczka.}}
 
\author{R.M. Weiner}

\date{
  Physics Department, University of Marburg, Marburg,
F.R
. Germany\\
E.mail: weiner@mailer.uni-marburg.de}

\maketitle

\begin{abstract}
It is shown that there exists an intimate relationship
between
Bose Einstein correlations and quantum field theory. On the
one hand several essential aspects of BEC cannot
be understood and even formulated without 
 second quantization. On the other hand BEC can serve as a
unique
tool
in the investigation of modern field theory and in
particular of the standard model. Some new developments on
this subject related to multiparticle production and 
squeezed states are also discussed.     
\end{abstract}

Bose Einstein correlations (BEC) are a topic of high current
interest in particle and nuclear physics. This interest has
been  motivated so far mainly by the fact that they offer a
unique possibility to explore the space-time dimensions of 
sources of particles and this is essential e.g. in the
search
for quark matter. However BEC can in principle offer much
more, namely insight into some fundamental aspects of
quantum mechanics, as well as the possibility to test
important aspects of modern particle physics. Historically
BEC came into being when Hanbury-Brown and Twiss invented in
the mid fifties the method of photon intensity
intereferomentry for the measurement of stellar dimensions
(the HBT method).  
 In 1959-1960 G.Goldhaber, 
S.Goldhaber,W.Lee and A.Pais discovered 
that identical charged pions produced in $\bar p-p$ annihilation
are correlated (the GGLP effect). Both the HBT and the GGLP 
effects are based on Bose-Einstein correlations.
Subsequently also 
Fermi-Dirac correlations for nucleons were observed.

Loosely
speaking
both Fermi-Dirac and Bose-Einstein correlations can be viewed as a consequence of the
symmetry (antisymmetry) properties of the wave function
with respect to permutation of two identical particles
with integer (half-integer) spin and are thus intrinsic
quantum phenomena.  At a higher level, these symmetry properties
of identical particles are expressed by the commutation relations
of the creation and annihilation operators of
particles in the second quantisation (quantum field theory).
The quantum field approach is the more general approach 
as it contains the possibility to deal with creation and 
annihilation of particles and certain correlation phenomena
like the correlation between particles and antiparticles can
be properly described only within this formalism.
Moreover, at high energies, because of the large number of particles
produced, not all particles can be detected in a given
reaction and therefore one measures usually only inclusive 
cross sections. For these reactions the wave function
formalism
is impracticable.
Furthermore, as pointed out quite recently \cite{Sj} BEC may
play
an important part in the test of the standard model and
in particular in the search for the Higgs particle, because
they may affect the W mass.
Last but not least BEC can serve for the
determination of one of the most characteristic properties of
systems made of identical bosons and which is responsable
for the phenomenon of {\em lasing} in quantum optics 
namely  quantum statistical
coherence. This feature is also not accessible to a
theoretical
treatment except in field theory.

\subsection*{BEC, coherent states, and the density matrix }
 
To realise the significance of this topic it is enough to
mention that some of the most important developments in 
particle physics
of the last 25 years including the ``standard model", 
are based on spontaneously broken
symmetries which imply coherent states. Moreover certain
classes of field theories admit classical fields as
solutions
(e.g. solitons) and any classical field is a coherent state. 
However there is so far no direct experimental evidence 
for these coherent states. On the other hand it is well known from
quantum optics that BEC depend on the amount of coherence
\footnote {In the last decade the conjugated quantity to
amount of coherence, i.e. the amount of {\em chaoticity} 
or just chaoticity,
as introduced by the author in Proceedings of LESIP II, Hadronic
Matter in Collision, World Scientific 1986, Eds. P.
Carruthers and D. Strottman, page 106 is more often used.
 It is perhaps
amusing to mention that the term chaoticity was proposed 
for the first time in a paper by Fowler, Friedlander, 
Weiner and Wilk submitted to Physical Review Letters, but
the editor of this journal objected to this word,  
 and we had to replace it by ``measure of chaos"
(cf. Fowler, Friedlander, Weiner and Wilk in Phys. Rev. Lett.   
57(1986) 2119). Since then, however, times, preconceptions
and editors have changed...At present chaoticity is a universally
accepted and used term.}     
in a very characteristic way 
 and therefore one hopes to obtain
information
about coherence from boson interferometry.

This dependence of BEC on coherence is a particular case
of the fact that any probability or cross section in
quantum theory is an expectation value and thus depends
 on the state of the system. 
This is demonstrated explicitely in the quantum
field
theoretical treatment of BEC \cite{Leo}. In general
the state of the system is described by the density matrix,
which in principle is determined by the theory. 
For hadron multiproduction this theory is quantum
chromodynamics and for processes involving multiphoton
production this theory is
quantum electrodynamics. However in both cases the use of
the fundamental theory is unpractical because of the
complexity of the many body problem. That is why one uses
in both cases phenomenological approaches. 
It is interesting to mention that for photonic processes
the invention of intensity interferometry by Hanbury-Brown
and Twiss led to the development of a new branch of optics,
namely quantum optics, with laser physics being one of its
major applications. The experience gained in this domain
has been instrumental in the analogous problem of hadron
multiparticle production and amounts essentially to
postulating
the form of the density matrix in the {\em coherent state}
representation.

\subsection*{ BEC and the notion of identical particles}
There are some aspects of principles of quantum mechanics
involved in the study of BEC, which have been discovered
more recently and which are related to the very concept
of {\em identity} of particles. It is well known that the 
principle of identity of particles is part of
the fundamental postulates of quantum mechanics and states
essentially that elementary particles are indistinguishable.  
The question what means {\em identical} has not been
raised until recently,
since it had been considered that the answer to it was
obvious. This situation has changed 
when it was discovered within the classical current formalism
 (cf.\cite{apw2}) that
 there exists a difference between BEC of
neutral and charged pions.
While the maximum value of the second order correlation
function for charged pions is 2, that for neutral pions
is 3. To realise in simpler terms the meaning of this, it is
useful to recall that Bose Einstein correlations imply in
general a
bunching of identical particles and the larger the intercept,
 the stronger this bunching is. Non identical bosons on
the
other hand do not show this bunching. Thus the bunching
phenomenon can be considered as a signal of
the
identity of particles. The fact that some identical bosons
are more bunched than others implies then in some sense
 that there exists a ``hierarchy" of identity as if some
identical particles would be more identical than others.
Another aspect of this phenomenon is the fact that the
amount
of bunching is a manifestation of the state of the system.
Thus a conventional coherent state has no bunching (similar to the
situation met with non identical particles), a chaotic state
has bunching and a squeezed coherent state can have 
any amount of bunching including negative values, i.e.
antibunching (cf.e.g.\cite{walls}).

An even more striking aspect of this effect is the fact
that there exists a quantum statistical correlation between
positive and negative pions \cite{apw2}, although these particles are
non identical in the ususal sense of the word.
These results appeared so surprising at the moment when they
were obtained that some people could not believe them 
and attempts were made to disprove them.
The reason for this reaction of the scientific community
lies in the fact that the naive wave function approach to BEC
was (is?) still deeply rooted and there is no obvious way
to obtain these surprising effects within this approach.
Quantum field theory i.e. second quantization of which
the clasical current formalism is a particular case, is 
the natural frame for the derivation of these new effects.
 At present there exist at least
three other derivations of these 
effects \cite{Bowler} 
and they 
constitute a definite challenge for
experimentalists. The fundamental importance of these
effects is such that their experimental observation will 
compensate by far
the efforts necessary for their detection (these effects 
are quite small and
necessitate high statistics to be seen in experiment).
It is important to emphasize that the quantum statistical
corelation between positive and negative pions mentioned
above is just a 
particular case of quantum statstical correlations
 between particles
and antiparticles in general and is not restricted to
isospin one (e.g. such a correlation must
exist also between positive and negative kaons or W bosons.).

 The classical current formalism has been used widely in the
context of BEC  
because for this case there
exist exact solutions of the corresponding inhomogeneous field 
equations. On the other hand it has been clear that in particle
physics the currents are quantised. Much less clear was
how to  estimate
 the quantum corrections to the classical currents. 
Furthermore one might have wondered whether the 
new effects discussed above and derived within
the classical formalism 
 were not an artifact of this formalism. 
In ref.\cite{Leo} an answer to the above questions is given by
formulating explicitely a quantum field theory of BEC
in which the quantum nature of the currents is taken into
account.
Concretely this means considering the currents as operators
which implies proper ordering in the corresponding
expressions
for the physical quantities which are calculated.
Besides confirming fully the existence of the new effects
quoted above the quantum field theoretical treatment
presented in \cite{Leo} brings   
another surprise in this saga of BEC. It turns out that  
these very effects can be used in order to study experimentally
  the quantum
corrections to the classical current approximation.

Finally we will describe a most recent development on the
subject of BEC and quantum field theory related to squeezed
states.

\subsection*{BEC and squeezed states}
Besides
ordinary
coherent states used as the basis of the representation,
squeezed coherent states have been introduced, which 
are of major interest both from a theoretical point of view 
as well as because of their application potential.
As will be shown below BEC can serve as a tool for 
the detection of these squeezed states.

Coherent states are the nearest approximation to classical 
fields because they minimise the product of incertitudes in
the
Heisenberg indeterminacy relation. They are defined as 
 eigenstates of the one particle annihilation 
operator. Besides these {\em ordinary} coherent states there
exist however also {\em squeezed} coherent states which are
eigenstates of the two or more particles annihilation
operator (cf. e.g. \cite{walls}). These generalised coherent
states, which are a $U(1,1)$ group extension of ordinary
coherent states, have been for the last years in the center 
of interest of several branches of physics.  
While for ordinary coherent states the
fluctuations in momentum and coordinate are equal to
the corresponding zero-point vacuum fluctuations, 
squeezed
states 
   allow for even smaller incertitudes (in one canonical 
variable). Thus the
quantum limit can be ``beaten" and this 
is not only of fundamental  
interest, but
may have important
applications in communication technology and for
measurements
of very weak signals (gravitational waves e.g.).

Although the effect of squeezed states in BEC has been 
discussed in the
optical and particle physics literature for quite some time,
this discussion has been limited so far to the idealised
case of pure squeezed states and even for this case only the
value of the second order correlation function in the origin
was known (cf.e.g.\cite{vwb},
 \cite{kogan}).  
Another important issue related to squeezed states is the fact
that while in optics squeezed states have been obtained 
in the last
years in several experiments, in particle physics
this is apparently not the case.
In a recent paper \cite{AW} progress along these lines could be
reported. In particular it has been shown that 
squeezed states could be produced 
preferentially in ``sudden" nuclear and particle reactions.
and a derivation of 
 the second order Bose Einstein correlation 
function in the entire domain of its variables has been given
for the practically important case of a chaotic superposition of
squeezed states. Furthermore it has been shown 
 that by measuring BEC in ``sudden" reactions
important new information about the dispersion of the 
hadronic medium before it emits can be obtained.
I shall sketch below briefly these results.
 
Consider 
 a blob of 
hadronic matter (for which Shuryak \cite{shury}
proposed the name ``pion liquid") created
in particle collision which undergoes a sudden breakup 
into free pions. In other
words, the pionic system, having its specific ground state and elementary pionic
excitations (not coinciding exactly with the usual vacuum and free particles)
converts rapidly into free pions. In this case the single and higher order
inclusive cross section and the many-particle correlation functions will depend
on the spectrum of excitations in the pionic system. The importance of the form
of the spectrum of pionic excitations for 
multiparticle production was stressed in the same reference
by Shuryak.
I shall argue below that the above physical picture results in the production
of quantum squeezed states.

Let us consider the transition from a pionic ``liquid" to a free pion field in the
spirit of local parton-hadron duality, i.e. we conjecture a close correspondence
between particles (fields) in the two ``phases".
 At the moment of this transition one can postulate
the following relations between the generalized coordinate $Q$
and the generalized momentum $P$ of the field:
\ba
Q &=& \frac{1}{\sqrt{2E_b}} (b^+ + b) = \frac{1}{\sqrt{2E_a}} (a^+ + a)
\nonumber\\
P &=& i \sqrt{\frac{E_b}{2}} (b^+ - b) = i \sqrt{\frac{E_a}{2}} (a^+ - a) 
\label{eq:qp}
\ea
$a^+,a$ are the free field creation and annihilation operators and $b^+,b$ the
corresponding operators in the ``liquid". Eq. (\ref{eq:qp}) holds for each mode
$p$. Then we get immediately a connection between the $a$ and $b$ operators,
\ba
a &=& b \; cosh \; r \; + \; b^+ \; sinh \; r \; , \nonumber\\
a^+ &=& b \; sinh \; r \; + \; b^+ \; cosh \; r \; 
\label{eq:aa}
\ea
with
\be
r = r (\vec p) = \frac{1}{2} \, log \, (E_a/E_b) \, .
\label{eq:rr}
\ee

The transformation (2) is just the squeezing transformation \cite{walls} with
a momentum dependent squeezing parameter $r(\vec p)$ given by eq. (\ref{eq:rr})
and the coherent eigenstate $|\beta>_b$ of the $b$-operator is the squeezed
state $|\alpha,r>_a$ of the $a$-operator:
\be
|\beta>_b = |\alpha,r>_a 
\label{eq:bb}
\ee
where $\alpha$ and $\beta$ are related by the same transformation (2) as the $a$
and $b$ operators. This proves the above made statement.

In general the system may not be in a pure coherent 
or squeezed state and then a
statistical averaging has to be 
performed both with respect to the coherent as
well as for the squeezed states. 
Interestingly enough in the case of squeezed states this
apparently routine task raises a new question of principle.

In practice it is easier to express the
$a,a^+$-operators through the $b,b^+$-operators according to eq. (\ref{eq:aa})
and then perform the averaging over the coherent states $|\beta>_b$. Considering
charged
identical pions (complex valued field) we shall use the Glauber-Sudarshan
representation of the density matrix, and write the average value
of an operator $\hat{O}$ as 
\be
<\hat O (a,a^+) > = \prod \limits_{\vec p} \int d^2 \beta_k P \{ \beta(\vec p)
\} <\beta|\hat{O} \, (a(b,b^+),a^+(b,b^+))|\beta>_b
\label{eq:oo}
\ee
and assume a Gaussian form for the weight function $P\{\beta(\vec p)\}$. Due to
the linearity of the squeezing transformation (2) this form will hold also for the
$a,a^+$-operators.

The particle source will be characterized by a
primordial correlator determined by the
number density $n(p)$ and by a function $f(\vec{x})$ describing its
geometrical form, see
refs. \cite{aw,apw}. To make contact with the previous results of \cite{aw,apw}
we note that the radius of the source $R$ enters the function $f$ and the
correlation length $L$ appears in $n(p)$. For simplicity we shall not consider
the time dependence here and take the form function $f(\vec x)$ to be dependent
only on the space coordinates. 

And now we arrive at a new
surprise:
the direct substitution of the transformation (2) into eq. (\ref{eq:oo}) leads
to undefined (divergent) expressions of the form $\delta(0)$
when one tries to perform the normal
ordering of $b,b^+$-operators (the last is necessary to use the coherent state
representation of eq. (\ref{eq:oo})). This situation 
can be avoided by introducing  new
creation and annihilation operators which are non-zero only inside the volume of
the particle source,
\be
\tilde a(\vec x) = a (\vec x) f (\vec x) \quad , \quad \tilde{a}^+(\vec x) =
a^+ (\vec x) f (\vec x) \; ,
\label{eq:aa2}
\ee
or, for Fourier transformed quantities,
\be
\tilde a (\vec p) = \int \frac{d^3k}{(2\pi)^3} a(\vec k) f(\vec k - \vec p)
\quad , \quad \tilde{a}^+ (\vec p) = \int \frac{d^3k}{(2\pi)^3} a^+(\vec k)
f(\vec p - \vec k)
\label{eq:aa3}
\ee
with standard commutation relations
\be
\left[ a(\vec{p}_1),a^+ (\vec{p}_2) \right] = (2\pi)^3 \cdot
\delta^3(\vec{p}_1-\vec{p}_2) \, .
\label{eq:aa4}
\ee

Then the equal momentum commutators of the modified operators are finite. For
example:
\be
[\tilde a (\vec p),\tilde{a}^+ (\vec p)] = \int \frac{d^3k}{(2\pi)^3}
f(\vec p - \vec k)f(\vec k - \vec p) = \int d^3x |f|^2(\vec x) = V_{eff}
\label{eq:aa5}
\ee
being equal to an effective volume $V_{eff}$ of the particle source. 
While this finite size is quite natural in particle physics, it is not 
so in optics where the system is usually macroscopic.
Furthermore it is remarkable that this problem of finite
size
appears only with squeezed states and only when correlations
are considered. Thus in \cite{v},\cite{vwa},\cite{vwb} where ``thermal" squeezed
states were introduced and applied to multiplicity
distributions and their moments (these are the {\em integrals}
 of
correlation functions) this did not happen..

With the smoothed operators $\tilde a(\vec p),\tilde{a}^+(\vec p)$ substituted
into eq. (\ref{eq:oo}) the form of the source is already taken into account and
the remaining statistical averaging may be performed in the same way as for
an infinite medium. 

Now the evaluation of the averaged matrix elements is straightforward.
Substituting eqs. (\ref{eq:aa3}) and (\ref{eq:aa}) into eq. (\ref{eq:oo}) and
performing the Gaussian averaging over coherent states $|\beta>_b$ we get the
single-particle inclusive density in the form:
\ba
\rho_1(\vec p)  &=& \frac{(2\pi)^3}{\sigma} \frac{d\sigma^{in}}{d^3p} =
<\tilde{a}^+(\vec p) \tilde{a}(\vec p)> \nonumber\\
&=& \int \frac{d^3k}{(2\pi)^3} \left[ n_b(\vec k) cosh \, 2r(\vec k) + sinh^2 r
(\vec k) \right] f(\vec p - \vec k) f(\vec k - \vec p)
\label{eq:rr2}
\ea
where the function $f$ describes the effect of finite size of the particle
source and $n_b(\vec k)$
given by the equation
\be
<\beta^*(\vec{k})\beta(\vec{k}')> = (2\pi)^3 \delta^3 (\vec{k}-\vec{k}')
n_b(\vec{k})
\label{eq:new}
\ee
represents the density of pionic ``quasiparticles"
($b$-quanta) (in particular, for a thermal source the
function $n_b(\vec k)$ is the usual Planck distribution function).

The squeezed state effect is reflected in eq.(\ref{eq:rr2}) in the factor
$cosh2r$ multiplying the primary pionic density 
$n_b(\vec k)$ and in the term $sinh^2r$ representing the ground state
contribution. That is the final state pions are produced even if the pions in
the pionic source are absent (zero temperature), just due to the decay of the
squeezed vacuum state. According to eq. (\ref{eq:rr2}), the single particle
density may be strongly enhanced in the presence of squeezed states if the
squeezing parameter $r(\vec p)$ is large enough. 

We consider now the two-particle inclusive density
\be
\rho_2(\vec{p}_1,\vec{p}_2) = \frac{(2\pi)^6}{\sigma} \cdot \frac{d \sigma}
{d^3p_1 \cdot d^3p_2} = <\tilde{a}^+(\vec{p}_1) \tilde{a}^+(\vec{p}_2)
\tilde{a}(\vec{p}_1) \tilde{a}(\vec{p}_2)>
\label{eq:rr3}
\ee
in the presence of squeezed states. With the finite size cut off the
two-particle density is calculated in
the same way as the single-particle density. Using Gaussian averaging 
one gets the simple expression
\be
\rho_2(\vec{p}_1,\vec{p}_2) = <\tilde{a}^+(\vec{p}_1) \tilde{a}(\vec{p}_1)> 
<\tilde{a}^+(\vec{p}_2) \tilde{a}(\vec{p}_2)> +
|<\tilde{a}^+(\vec{p}_1) \tilde{a}(\vec{p}_2)>|^2 + |<\tilde{a}(\vec{p}_1)
\tilde{a}(\vec{p}_2)>|^2
\label{eq:rr4}
\ee
with
\ba
<\tilde{a}^+(\vec{p}_1) \tilde{a}(\vec{p}_2)> &=&
\int \frac{d^3k}{(2\pi)^3} \left[ n_b(\vec k) cosh \, 2r(\vec k) + sinh^2 r
(\vec k) \right] f(\vec{p}_1 - \vec k) f(\vec k - \vec{p}_2), \nonumber\\
<\tilde{a}(\vec{p}_1) \tilde{a}(\vec{p}_2)> &=&
\int \frac{d^3k}{(2\pi)^3} \left[ n_b(\vec k) + \frac{1}{2} \right] sinh \, 2r
(\vec k) f(\vec{k}_1 - \vec{p}_1) f(\vec k - \vec{p}_2) \, ,
\label{eq:aa6}
\ea
   
The first term in the right hand side of eq. (\ref{eq:rr4}) is the product of
single-particle densities $\rho_1(\vec{p}_1) \rho_1(\vec{p}_2)$, the second term
is the exchange contribution characteristic for Bose-Einstein correlations
modified by the squeezing factor $r$ (for $r = 0$ it coincides with the
usual BEC).
The third term arises only in the presence of squeezed states (it
vanishes for $r = 0$). This last contribution differs from the ``surprising"
terms in the two-particle correlation function discussed in refs.
\cite{apw,apw2}, which are absent in the case of charged identical pions under
consideration and which have another dependence on momenta
$\vec{p}_1,\vec{p}_2$, being maximal at $\vec{p}_1 + \vec{p}_2 = 0$, and not at
$\vec{p}_1 - \vec{p}_2 = 0$ as is the case for all terms in eq. (\ref{eq:rr4}).

As one can see from eqs. (\ref{eq:rr2}), (\ref{eq:rr4}),(\ref{eq:aa6}) the
second order correlation function
\be
C_2(\vec{p}_1,\vec{p}_2) = \rho_2(\vec{p}_1,\vec{p}_2)/[\rho_1(\vec{p}_1)\rho_1
(\vec{p}_2) ]\, .
\label{eq:rr5}
\ee
is enhanced due to the presence
of the third term in the right hand side of eq. (\ref{eq:rr4}) and in general
the value of the ratio (\ref{eq:rr5}) is arbitrarily large. In particular, for
$n_b(\vec k) = 0$ (that is for cold pionic matter when particle production is
the result of the squeezed vacuum decay) and for small values of the
squeezing parameter
$r(\vec{k})$, one may have $C_2 >> 1$ . For $r(\vec{k}) \sim 1$ and $\vec{p}_1 \cong \vec{p}_2$ the ratio
(\ref{eq:rr5}) is close to three. We call this effect 
``overbunching" to distinguish it from conventional Bose
Einstein correlation where a bunching effect occurs, too,
but
where the intercept $C_2(p,p)$ does not exceed the value of
two.

Possible applications of the rapid transition mechanism 
discussed above could include the explosion of a hadronic fireball after a phase transition 
from quark-gluon plasma \cite{gyul} and annihilation.
\footnote{
 An overbunching effect similar to that discussed above has
been seen in annihilation reactions \cite{adler} but 
was interpreted in \cite{song} in terms of resonance 
production. 

The present approach in terms of squeezed states should be
viewed as an alternative or a supplementary source of
overbunching. It has the advantage that it leaves
transparent the relation between BEC and the space-time 
characteristics of the source and 
it
could perhaps also
explain why overbunching may have been seen so far only in 
annihilation.
Indeed
our derivation of the squeezing effect assumes a rapid 
transition and it has been argued recently by Amado et al.
\cite{amado}                that
annihilation is a fast process.} 

{\bf Acknowledgements}\\
The author is indebted to R.Raczka and G.Wilk for the kind hospitality
extended to him in Poland and to M.Pawlowski and the other organisators of
this interesting meeting for the fine job they did.

Instructive
discussions with A. Capella, J. Letessier, U. Ornik, M.
Pl\"umer, 
Y. Sinyukov
A.Tounsi and A. Vourdas are gratefully acknowledged.

\end{document}